\documentstyle[11pt,epsfig]{article}
\def\appendix{\par
 \setcounter{section}{0}
 \setcounter{subsection}{0}
 \def\thesection{Appendix \Alph{section}}
 \def\theequation{\Alph{section}.\arabic{equation}}
 \setcounter{equation}{0}}

\def\a{\alpha}

\def\le{\left(}
\def\ri{\right)}

\def\no{\nonumber}
\def\G{\Gamma}
\def\rar{\rightarrow}

\def\e{\epsilon}
\def\f12{\frac{1}{2}}

\def\pd{\partial}

\def\L{\lambda}
\newcommand{\Li}{\mathop{\mathrm{Li}}\nolimits}

\begin{document}
\begin{titlepage}
%\flushright{}
\vskip 2cm
\begin{center}
{\Large \bf  New four-dimensional integrals by Mellin-Barnes transform}\\
\vskip 1cm  
Pedro Allendes$^{a},$ Natanael Guerrero$^{b},$ Igor Kondrashuk$^{b},$  Eduardo A. Notte Cuello$^{c}$\\
\vskip 5mm  
{\it  (a) Departamento de F\'\i sica, Universidad de Concepci\'on, Casilla 160-C, Concepci\'on, Chile} \\
{\it  (b) Departamento de Ciencias B\'asicas,  Universidad del B\'\i o-B\'\i o, \\ 
            Campus Fernando May, Casilla 447, Chill\'an, Chile} \\
{\it  (c)  Departamento de Matem\'aticas, Facultad de Ciencias, Universidad de La Serena, \\ 
               Benavente 980, La Serena, Chile}
\end{center}
\vskip 20mm

\begin{abstract}
This paper is devoted to the calculation by Mellin-Barnes transform of a especial class of
integrals. It contains double integrals in the position space in $d = 4-2\e$ dimensions, where $\e$
is parameter of dimensional regularization. These integrals contribute to the effective action
of the ${\cal N} = 4$ supersymmetric Yang-Mills theory. The integrand is a fraction in which the
numerator is a logarithm of ratio of spacetime intervals, and the denominator is the product of 
powers of spacetime intervals. According to the method developed in the  previous papers, in order
to make use of the uniqueness technique for one of two integrations, we shift exponents
in powers in the denominator of integrands by some multiples of $\e$. As the next step, the
second integration in the position space is done by Mellin-Barnes transform. For normalizing
procedure, we reproduce first the known result obtained earlier by Gegenbauer polynomial
technique. Then, we make another shift of exponents in powers in the denominator to create
the logarithm in the numerator as the derivative with respect to the shift parameter $\delta.$
We show that the technique of work with the contour of the integral modified in this way
by using Mellin-Barnes transform repeats the technique of work with the contour of the
integral without such a modification. In particular, all the operations with a shift of contour
of integration over complex variables of two-fold Mellin-Barnes transform are the same as
before the $\delta$ modification of indices, and even the poles of residues coincide. This confirms
the observation made in the previous papers that in the position space all the Green function 
of ${\cal N} = 4$ supersymmetric Yang-Mills theory can be expressed in terms of UD functions.
\end{abstract}
\end{titlepage}

\section{Introduction}

The method of uniqueness was a big jump in calculating of Green functions in the massless quantum field theories 
\cite{Unique,Vasiliev:1981dg,Vasil, Kazakov:1984bw}. It can successfully be used in a combination with 
a set of other powerful methods to calculate integrals, for example with Gegenbauer polynomial technique (GPT) of 
Refs.~\cite{Tkachov:1981wb,Chetyrkin:1981qh,Chetyrkin:1980pr,Celmaster:1980ji,Terrano:1980af,Lampe:1982av,Kotikov:1995cw} 
or with Mellin-Barnes (MB) transformation \cite{Smirnov:2004ym,Boos:1990rg}. Certain modifications of uniqueness method were developed  
and applied to so called one-step deviating from the unique objects \cite{Usyukina:1983gj}. 
The unique objects are unique triangle and unique vertex \cite{Usyukina:1983gj}. In Ref.~\cite{Usyukina:1983gj} it was observed 
that the number of rungs in a propagator-type diagram of a particular topology can be reduced by one in a special limit. 
Later, it has been realized that the number of rungs can be reduced by one in a triangle ladder-like vertex without that special 
limit \cite{Belokurov:1983km,Usyukina:1991cp}. The approach of 
Ref.~\cite{Usyukina:1983gj} helped to calculate a special class of propagator-type diagrams for an arbitrary number of loops 
in Ref.~\cite{Belokurov:1983km}.  The same approach was applied to calculation of triangle ladder vertex diagram 
with an arbitrary number of rungs and remarkable recursive relations between two neighbour-rung results were derived
\cite{Usyukina:1992jd,Usyukina:1993ch}.  In such a way an infinite series of UD functions 
$\Phi^{(n)}\le x,y\ri $ of two real variables $x,y$ was generated \cite{Usyukina:1992jd,Usyukina:1993ch}, $n$ is a number of the 
function in this series.

In the present paper we argue that the three-point scalar integrals in the position space in the massless four-dimensional quantum field 
theories with certain powers of spacetime distances in the denominators can be reduced to the UD functions. 
Such integrals appear usually in a course of evaluation  of the three-point Green functions with several internal vertices 
in the position space. After evaluating the first of the integrations (or in other words, after making integration in the first of 
the internal vertices of a graph), in the next integrations we have the powers of 
spacetime intervals in the denominators changed from the initial values to integer numbers shifted by some multipliers 
of $\e,$ where $\e$ is a parameter of dimensional regularization, $d=4-2\e.$ With shifting these powers of spacetime distances  
in the denominator by a parameter $\delta$ we can study integrals with powers of logarithms in the numerator of the integrand.

It is necessary to mention that any three-point integration in the position space in the massless theory with any powers 
of spacetime intervals in the denominator can be represented in terms of Appell's hypergeometric function \cite{Davydychev:1992xr}. 
However, in a particular case of the integrals considered in the present paper which can be decomposed in Laurent 
series of the deviations parameters $\e$ and $\delta,$ the Appell's hypergeometric function is reduced to a combination of 
the UD functions.  In Ref.~\cite{Davydychev:1992xr} the Appell's function appears as the result of the residue calculation  
via Mellin-Barnes transform. This powerful method allows to perform complicate multiloop calculations 
in the quantum field theory \cite{Bern:2005iz}. In the present paper we apply  MB transform too. The method that we propose is based on 
the manipulations with contour of integration in the complex plane of MB parameters 
together with adding and subtracting certain products of the spacetime distances with the UD functions in the MB representation.
It can be applied to a more general case of the multipoint Green functions or amplitudes, for example to the recent 
Wilson line calculation \cite{DelDuca:2009au}.

One of the motivations for the investigation developed in this paper was  to calculate an  auxiliary vertex, 
which in  ${\cal N} = 4$ supersymmetric Yang-Mills theory does not depend on any scale, ultraviolet or infrared. 
This is $Lcc$ vertex in which the auxiliary field $L$ couples to two (self-adjoint) Faddeev-Popov ghost 
fields $c.$  It is superficially convergent in the Landau gauge in the component formalism for any renormalizable four-dimensional gauge 
theory. Formally, the superficial convergence of this vertex  holds to all orders of perturbation theory due to the 
antighost equation \cite{Blasi:1990xz}.  The superficial convergence for this vertex 
is equivalent to finiteness of this vertex for all number of loops in  ${\cal N} = 4$ supersymmetric Yang-Mills theory 
\cite{Cvetic:2004kx,Kondrashuk:2004pu,Cvetic:2006kk}. This is a consequence of ST identity 
\cite{Slavnov:1972fg,Taylor:1971ff,Slavnov:1974dg,Faddeev:1980be,Lee:1973hb,Zinn-Justin:1974mc,Becchi:1974md,Tyutin:1975qk} 
and the explicit two-loop result has been found in Refs.~\cite{Cvetic:2006iu,Cvetic:2007fp,Cvetic:2007ds}.
All the poles in $\e$ disappear in all number of loops for this vertex. However, it cannot be analysed by the methods of conformal field 
theory since the auxiliary field $L$ does not propagate \footnote{In contrary, three-point function of dressed mean gluon of Refs. 
\cite{Cvetic:2004kx,Kondrashuk:2004pu,Cvetic:2006kk} can be fixed by conformal symmetry  in analogy to 
\cite{Mitra:2008yr,Mitra:2008pw,Mitra:2009zm,Freedman:1998tz,Erdmenger:1996yc}, however, to find four-point off-shell 
correlator by conformal symmetry is impossible since an arbitrariness arises.}. In the nonsupersymmetric theories this vertex is 
not finite and a calculation of the anomalous dimension of operator $cc$ has been performed in \cite{Dudal:2003pe,Dudal:2003np}.

It is possible to find some hints for the relations derived in this paper studying the all-order finite $Lcc$ vertex. As it is 
described in detail in Section 3 of the present paper we can calculate a part of the three-loop correction to this vertex 
in two distinct ways. On one side, we can use the method of Ref. \cite{Cvetic:2007ds} and the result is the UD functions, 
logarithms and powers of spacetime intervals. On the other side, we can integrate the two-loop result of Ref. \cite{Cvetic:2007fp}
in the way as it is done usually in Bethe-Salpeter equation with one new gluon propagator  between external legs of 
the ghost fields for this vertex. The two-loop result contains logarithms of ratios of the spacetime distances between 
the three external points of this correlator, and in turn it is the integrand of the three-loop contribution 
done in the second way. Thus, the integration of the two-loop result should produce the UD functions at most. 
Such a type of integrands with the logarithms in the denominators is of the type that we consider 
in the present  paper. The all-order structure of this vertex  can be fixed by solving Bethe-Salpeter (BS) 
equation \cite{Kondrashuk:2008xq}. A solution to that equation would give the recursive coefficients in front 
of the UD functions which are the basis for decomposition of the $Lcc$ correlator 
in higher orders \cite{Kondrashuk:2008xq} of the perturbation theory.

The paper is organized as follows. In Section 2 we review the notation and previous results. 
In Section 3 the analysis and description of the integral structure of the vertex is performed. Section 4 is devoted to the reproducing 
via the Mellin-Barnes (MB) technique the result for $J(\L,\L,2\L)$ derived early by Gegenbauer polynomial technique. 
In Section 5 the integral $J(\L-\delta,\L,2\L+\delta)$ is taken via Mellin-Barnes (MB) technique in which $\delta$ is a small 
positive variable and $\L = 1-\e.$ In four Appendices some technical points are described. All the calculations are done in
the position space.

\section{Notation, technique and previous results}

\noindent In Refs.~\cite{Cvetic:2006iu,Cvetic:2007fp,Cvetic:2007ds} uniqueness method \cite{Kazakov:1984bw} and Gegenbauer polynomial 
technique (GPT) \cite{Kotikov:1995cw} were used in order to calculate correlators.  The uniqueness method contains several 
rules to integrate  massless chains and vertices algebraically, i.e. without a direct calculation of $d$-space integrals. 
We assume the notation of Ref. \cite{Cvetic:2006iu},  where $[Ny]= (x_N - y)^2$ and analogously for $[Nz]$ and $[yz],$ 
that is, $N=1,2,3$ stands for $x_N=x_1,x_2,x_3,$ respectively. The rules and notation are  following:

\begin{itemize}

\item The results for chains $J(\alpha_1,\alpha_2):$
\begin{eqnarray*}
J(\a_1,\a_2)  \equiv \int Dx \frac{1}{[x1]^{\a_1} [x2]^{\a_2}} = \frac{A(\a_1,\a_2,d-\a_1-\a_2)}{[12]^{\a_1 + \a_2 - d/2}}, \\
A(\a_1,\a_2,\a_3)= a(\a_1)a(\a_2)a(\a_3).
\end{eqnarray*}
We have introduced a new $d$-dimensional measure $Dx \equiv \pi^{-\frac{d}{2}}d^d x.$

\item Uniqueness method: if  $\a_1 + \a_2 + \a_3 = d$, then 
\begin{eqnarray*}
J(\a_1,\a_2,\a_3)  \equiv \int Dx \frac{1}{[x1]^{\a_1} [x2]^{\a_2} [x3]^{\a_3}} =  
\frac{A(\a_1,\a_2,\a_3)}{[12]^{d/2 - \a_3}[13]^{d/2 - \a_2}[23]^{d/2 - \a_1}}  
\end{eqnarray*}

\item  The result for the first UD function  \cite{Davydychev:1992xr} is  
\begin{eqnarray*}
J(1,1,1) = \frac{1}{[31]} \Phi^{(1)}\le \frac{[12]}{[31]},\frac{[23]}{[31]}\ri \\
=  \frac{2}{B} \left[ \zeta(2) - \Li_2 \left(\frac{[23]+[12]-[13]-B}{2[23]} \right) - \Li_2 \left(\frac{[23]+[13]-[12]-B}{2[23]} \right) 
\right. \\
\left. + \ln \left(\frac{[23]+[12]-[13]-B}{2[23]} \right) \ln \left(\frac{[23]+[13]-[12]-B}{2[23]} \right) \right.\\ 
\left. - \frac{1}{2} \ln \left(\frac{[12]}{[23]} \right) \ln \left(\frac{[13]}{[23]} \right) \right], \\
 B^2 = ([12]-[13])^2-2([12]+[13])[23]+[23]^2. 
\end{eqnarray*}

\end{itemize}
The ghost field interacts  with the gauge field only. This means that one proper diagram is possible at one-loop level 
for the auxiliary $Lcc$ vertex which contributes to the effective action as \cite{Cvetic:2007fp} 
\begin{eqnarray*}
\int~d^4x_1d^4x_2d^4x_3 \frac{i g^2 N}{2^{8}\pi^6} f^{abc} L^a(x_1)c^b(x_2)c^c(x_3) ~ V^{(1)} (x_1,x_2,x_3), \\
V^{(1)}(x_1,x_2,x_3)    \\
= \left[\frac{-1}{[12]^2[23]^2} + \frac{2}{[12]^2[31]^2}  + \frac{-1}{[23]^2[31]^2} \right.\\ 
+ \left.\frac{-1}{[12][23][31]^2} + \frac{2}{[12][23]^2[31]} +  \frac{-1}{[12]^2[23][31]}  \right]. 
\end{eqnarray*}
Two-loop planar correction  to the $Lcc$ vertex can be represented as combination of 
five diagrams of Ref.\cite{Cvetic:2007fp}, and the contribution of these diagrams to the effective action  is 
\begin{eqnarray}
\int~d^4x_1d^4x_2d^4x_3 \frac{i g^4 N^2}{2^{15}\pi^8} f^{abc} L^a(x_1)c^b(x_2)c^c(x_3) ~ V^{(2)} (x_1,x_2,x_3), \no\\
V^{(2)}(x_1,x_2,x_3)  =   \no\\
\left[\frac{29/3}{[12]^2[23]^2} + \frac{32/3}{[12]^2[31]^2}  + \frac{29/3}{[23]^2[31]^2}  + \frac{-52/3}{[12][23][31]^2} 
+ \frac{-40/3}{[12][23]^2[31]} +  \frac{-52/3}{[12]^2[23][31]}  \right] \no\\
+ \left[\frac{1}{[12]^2[23]} + \frac{1}{[23][31]^2} + \frac{-13}{[12][31]^2} + \frac{-13}{[12]^2[31]}  
+ \frac{-2[12]}{[23]^2[31]^2} \right.\no\\ 
\left. +  \frac{-2[31]}{[12]^2[23]^2}  + \frac{12[23]}{[12]^2[31]^2}  \right]J(1,1,1) \no\\
+ \left[\frac{-2}{[12]^2[23]^2} + \frac{-6}{[12]^2[31]^2} + \frac{18}{[23]^2[31]^2} + \frac{10}{[12][23][31]^2} \right.\no\\
\left.   + \frac{-14}{[12][23]^2[31]} + \frac{6}{[12]^2[23][31]}  \right]\ln[12] \no\\
+ \left[ \frac{-16}{[12]^2[23]^2} + \frac{12}{[12]^2[31]^2} + \frac{-16}{[23]^2[31]^2} + \frac{-16}{[12][23][31]^2} \right.\no\\  
\left. + \frac{28}{[12][23]^2[31]}  + \frac{-16}{[12]^2[23][31]}\right]\ln[23] \no\\
+ \left[\frac{18}{[12]^2[23]^2} + \frac{-6}{[12]^2[31]^2} + \frac{-2}{[23]^2[31]^2} + \frac{6}{[12][23][31]^2} \right.\no\\  
\left. + \frac{-14}{[12][23]^2[31]} + \frac{10}{[12]^2[23][31]}  \right]\ln[31].  \label{two-loop}
\end{eqnarray}

\section{Logarithmic integral}

We need to calculate a contribution to the effective action proportional to
\begin{eqnarray*}
\int~d^4x_1d^4x_2d^4x_3~ f^{abc} L^a(x_1)c^b(x_2)c^c(x_3) ~ V^{(3)} (x_1,x_2,x_3),
\end{eqnarray*}
in which the integral expression $V^{(3)} (x_1,x_2,x_3)$ is obtained as 
\begin{eqnarray}
V^{(3)} (x_1,x_2,x_3) = \Pi_{\mu\nu}(23)\pd^{(2)}_\mu\pd^{(3)}_\nu\int~d^4y~d^4z~\frac{V^{(2)} (x_1,y,z)}{[y2][z3]}, \label{V3}
\end{eqnarray}
where $V^{(2)}$ is the two-loop contribution (\ref{two-loop}). Eq.(\ref{V3}) is not the complete  planar contribution to the 
effective action at the three loop level. This is a part of it. The complete contribution includes all the graphs generated 
by Wick theorem. However, this is an important part and knowledge 
of how to calculate this integral allows us to derive  solution for the Bethe-Salpeter equation for this vertex and  to 
find coefficients in front of the UD functions, logarithms and powers of spacetime intervals in a recursive way. 
The present paper is devoted to analysis of the integrand structure in Eq.(\ref{V3}).

Let us consider the term in  $V^{(2)} (x_1,y,z)$  that contains logarithmic function  of ratio of the space-time intervals of the form
\begin{eqnarray}
\int~d^4y~d^4z~\frac{\ln{\le[yz]/[1y]\ri}}{[1y]^2[1z][yz][y2][z3]}. \label{obj} 
\end{eqnarray}
The integrand contains the logarithm in the numerator and the powers in the denominator. 
It is not straightforward that integrals of this 
type are reduced to the UD functions. There is a motivation for this. A calculation 
of the three loop ladder diagram of $Lcc$ vertex (1PI contribution with three gluon ``horizontal'' lines) can be done 
according to the rules of Ref.\cite{Cvetic:2007ds,Cvetic:2007fp}, however the result of this integration is the UD functions, 
logarithms of ratios of the intervals, and powers of spacetime intervals. On the other side, this ladder also contains 
contribution of this type  of integrands if the two-loop result of Ref.\cite{Cvetic:2006iu} for the ladder diagram is under 
integration. Since the first method does not give a new structure, the same conclusion should be true for the second way. 
Thus, it is expected that the integral (\ref{obj}) can be expressed in terms of the UD functions too.

The integral (\ref{obj}) is safe in the infrared region but is divergent in the ultraviolet region of integration. 
Thus, the dimensional regularization is needed. However, the total result for Eq.(\ref{V3}) should be finite in both the 
potentially dangerous regions of integration, ultraviolet and infrared. This means that all these singularities, 
realized as poles in $\e,$ should disappear in the final result for (\ref{V3}), and it makes possible to take the limit $\e \rar 0.$ 
The reason is that the integration of the first three two-loop planar diagram of Ref.\cite{Cvetic:2007fp}
with one new horizontal line as in Eq.(\ref{V3}) should be finite, since each of three contributions (now they are subgraphs) 
is finite, and every diagram contributing to this auxiliary vertex is finite superficially.

We use the technical trick of Ref.\cite{Cvetic:2006iu} shifting the powers in the denominators by multipliers of $\e$
in order  to remove one of the integrations in the position space by the uniqueness technique. Thus, we should find the integral
\begin{eqnarray}
I \equiv \int~Dy~Dz~\frac{\ln{\le[yz]/[1y]\ri}}{[1y]^{2-2\e}[1z]^{1-2\e}[yz][y2][z3]}. \label{obj-mod} 
\end{eqnarray} 
We can generate the logarithmic function in the numerator as 
\begin{eqnarray}
I = \le\frac{d}{d\delta}\int~Dy~Dz~\frac{1}{[1y]^{2-2\e+\delta}[1z]^{1-2\e}[yz]^{1 -\delta}[y2][z3]}\ri_{|_{\delta = 0}}. 
\label{obj-del} 
\end{eqnarray} 
The result is reduced to the analysis of the integral $F,$ 
\begin{eqnarray}
F \equiv \int~Dy~Dz~\frac{1}{[1y]^{2-2\e+\delta}[1z]^{1-2\e}[yz]^{1 -\delta}[y2][z3]}.  \label{Int}
\end{eqnarray}
If $\delta = 0$ we reproduce the analysis of the paper \cite{Cvetic:2006iu}, in which the integration over $y$ can be done by 
the uniqueness technique. The same method can be applied here, 
\begin{eqnarray}
F = A(2-2\e+\delta,1-\delta,1) \frac{1}{[12]^{1-\e+\delta}} \int~Dz~\frac{1}{[1z]^{2-3\e}[z2]^{\e-\delta}[z3]} \no\\
=   A(2-2\e+\delta,1-\delta,1) \frac{1}{[12]^{1-\e+\delta}} J(2-3\e,\e-\delta,1)\no\\
=  \frac{[23]^{1-2\e+\delta}}{[12]^{1-\e+\delta}}
\frac{ A(2-2\e+\delta,1-\delta,1)A(2- 3\e,1+\e-\delta,1+\delta)}{A(1+\e-\delta,1-\e,2-2\e+\delta)} \no\\
\times J(1-\e-\delta,1-\e,2-2\e+\delta) \label{F}
\end{eqnarray}
The latter integral has been  transformed as
\begin{eqnarray*}
J(2-3\e,\e-\delta,1) = \frac{1}{[23]^{1-2\e+\delta}} J(2-3\e,\e - \delta,1) [23]^{1-2\e+\delta} \\ 
= \int Dx \frac{1}{[23]^{1-2\e+\delta} [x1]^{2-3\e} [x2]^{\e- \delta} [x3]}[23]^{1-2\e+\delta}    \\
= \frac{1}{A(1+\e-\delta,1-\e,2-2\e+\delta)}[23]^{1-2\e+\delta} \\
\times \int Dx  \frac{1}{[x1]^{2-3\e}} 
~ \int Dy \frac{1}{[yx]^{1+\e-\delta} [y2]^{1-\e} [y3]^{2-2\e+\delta}} \\
= [23]^{1-2\e+\delta} \frac{A(2- 3\e,1+\e-\delta,1+\delta)}{A(1+\e-\delta,1-\e,2-2\e+\delta)}
\int Dy \frac{1}{[y1]^{1-\e-\delta} [y2]^{1-\e} [y3]^{2-2\e+\delta}}   \\
= [23]^{1-2\e+\delta} \frac{A(2- 3\e,1+\e-\delta,1+\delta)}{A(1+\e-\delta,1-\e,2-2\e+\delta)}J(1-\e-\delta,1-\e,2-2\e+\delta).
\end{eqnarray*}

In Ref.\cite{Cvetic:2006iu} Gegenbauer polynomial technique (GPT) has been applied 
in order to take the integration in $J(1-\e-\delta,1-\e,2-2\e+\delta)$ in the case $\delta = 0.$ 
However, it is not justified in the present paper since $F$ is not reduced to the integral with 
two indices $1-\e$ of the three indices of three-point integration. This makes GPT inefficient for the 
case when  $\delta \ne 0.$ In the next section we will show that 
Eq.(\ref{F}) is expressed in terms of the UD functions by using Mellin-Barnes technique. In the work with MB transformation  we mainly 
follow the methods described in known textbook \cite{Smirnov:2004ym} and in Ref.\cite{Boos:1990rg}, transforming the contour of 
integration according to Cauchy theorem.

The shift (\ref{obj-mod}) must be done in the same manner for all the terms in Eq.(\ref{V3}). As to other terms that 
can appear in the integration in  Eq.(\ref{V3}), 
including the integrands with pure powers and the terms proportional to $J(1,1,1),$  they are reduced again to  
the terms proportional to the UD functions, or to the pure powers, or to the terms with logarithms. 
For the pure powers, the argument repeats the approach of the previous papers \cite{Cvetic:2006iu,Cvetic:2007fp,Cvetic:2007ds}, 
and for $J(1,1,1)-$terms the argument is  more difficult but still is a quite direct.

\section{Mellin-Barnes transform for $J(\L,\L,2\L)$}

The MB transform of the integral with three external points is 
\begin{eqnarray}
J(\nu_1,\nu_2,\nu_3) = \frac{1}{\Pi_{i}\G(\nu_i)\G(d-\Sigma_i \nu_i)}
\oint_C dz_2~dz_3 \frac{\le[31]/[23]\ri^{z_2} \le [12]/[23]\ri^{z_3}}{{[23]^{\Sigma \nu_i - d/2}}} 
\left\{\G \le -z_3 \ri \G \le -z_2 \ri \right.\no\\
\left.\G \le -z_2-\nu_1-\nu_3 + d/2 \ri \G \le -z_3-\nu_1-\nu_2 + d/2 \ri 
\G \le  \Sigma \nu_i - d/2 + z_3 + z_2 \ri \G \le z_2 + z_3  + \nu_1 \ri \right\}. \label{J-arb}
\end{eqnarray}
In the rest of the paper we use the notation 
\begin{eqnarray*}
x \equiv \frac{[31]}{[23]},~~~ y \equiv \frac{[12]}{[23]},~~~\L = 1 -\e.
\end{eqnarray*}
The contour of integration $C$ passes a bit on the left of the imaginary axis, separates left and right poles and should be closed 
to the left infinity or to the right infinity. Whether it should be closed to the left infinity or to the right infinity 
depends on the relative value of the space-time intervals between the three points $x_1,x_2,x_3.$ Supposing that $x<1,y<1$
we close the contour of integration in the complex plane to the right infinity.  It could be closed to the left 
infinity too (in the opposite case if $x>1,y>1$) but it makes more complicate to take the residues into account since the residues
in variables $z_2$ and $z_3$ are mixed in that case. Whether we have to close the contour to the right infinity or to 
the left infinity, the result should be the same function. We omit the factor $1/2\pi i$ that accompanies each integration over 
MB transform parameter. The inverse factor is generated in front of the residues.

It is known that this representation of the three-point integral can be derived by applying two-fold MB transform 
to the integral over Feynman parameters, producing the Euler beta functions \cite{Smirnov:2004ym}
after integrating these parameters. 
There is a difference with the representation used in Ref.\cite{Usyukina:1993ch}. The form of Ref.\cite{Usyukina:1993ch}
can be recovered from (\ref{J-arb}) by a cyclic redefinition. In the rest of the paper, we use representation (\ref{J-arb}).

In order to show that Eq.(\ref{Int}) can be expressed in terms of the UD functions, we show first how to recover  
Eq.(14) obtained in Ref. \cite{Cvetic:2006iu} by GPT, via the Mellin-Barnes transform. The same derivation will be significantly used  
with a little modification in the next section. The MB transform for $J(\L,\L,2\L)$ is a particular case of Eq.(\ref{J-arb}),
\begin{eqnarray}
J(\L,\L,2\L) = \frac{2\e}{\G^2(1-\e)\G(2-2\e)\G(1+2\e)} \oint_C dz_2~dz_3~ \frac{x^{z_2}~y^{z_3}}{[23]^{2-3\e}} 
\left\{\G \le -z_3 \ri \G \le -z_2 \ri \right.\no\\
\left.\G \le -z_2 -1 +2\e\ri \G \le -z_3 + \e \ri \G \le z_3 + z_2 + 2 -3\e \ri \G \le z_2 + z_3  + 1 - \e \ri \right\}.\label{J-part}
\end{eqnarray}

Our purpose is to represent $J(\L,\L,2\L)$ in terms of the first  UD function $J(1,1,1),$ which has the integral MB representation 
\begin{eqnarray}
J(1,1,1) = \oint_C dz_2~dz_3~ \frac{x^{z_2}~y^{z_3}}{[23]} \G^2 \le - z_3 \ri \G^2 \le - z_2 \ri \G^2 \le 1 + z_2 + z_3 \ri,
 \label{J-111}
\end{eqnarray}
by transforming the contour of integration in (\ref{J-part}) in order  to rectify the vertical part of contours.
What is the most important in the integral MB representation of $J(1,1,1)$ is that the contour $C$ 
in the complex planes $z_2$ and $z_3$ includes the vertical line a bit on the left of the imaginary axis 
in the both planes and the contour is closed to the right infinity. We choose the real part of $\e$ is positive, Re $\e > 0.$

By construction of the MB transform, contours of the MB integration separate left and right poles in complex 
planes of the parameters of the MB transformation. In case of Eq.(\ref{J-part}) the most left of the right poles is the pole 
$z_2 = -1 + 2\e$ in $z_2$ plane and the rest of the right poles are situated on the right half-planes of $z_2$ and $z_3$ complex planes. 
Thus, we have to separate from the set of the right residues in $z_2$ plane the residue at this pole  
(due to Cauchy theorem) in order to extract the structure corresponding to the vertical line of the integration 
contour $C,$ 
\begin{eqnarray}
J(\L,\L,2\L) = R + 2\e I_1 + o(\e),  \label{res} \\ 
R \equiv  \frac{2\e~\G \le 1-2\e \ri}{\G^2(1-\e)\G(2-2\e)\G(1+2\e)}\frac{x^{-1+2\e}}{[23]^{2-3\e}} \no\\
\times \oint_C dz_3~ y^{z_3}\left\{\G \le -z_3 \ri \G \le -z_3 + \e \ri \G \le z_3 + 1 - \e \ri \G \le z_3  + \e \ri\right\}, \no\\
I_1 \equiv \oint_C dz_2~dz_3~ \frac{x^{z_2}~y^{z_3}}{[23]^2} \left\{ \G^2 \le -z_3 \ri 
\G \le -z_2 \ri  \G^* \le -z_2 -1 \ri \G \le z_3 + z_2 + 2 \ri  \G \le z_2 + z_3  + 1\ri \right\}  \no
\end{eqnarray}
In integral $I_1$ we have removed the dependence on $\e$ in the integrand since the poles in $\e$ cannot appear in it, and moreover 
we do not need the term of the second order in $\e.$ Absence of poles in $\e$ follows by the observation that no effect of gluing the 
left and right poles presents in that integral after extracting the contribution of the pole  $z_2 = -1 + 2\e.$ 
Only gluing effect between poles of distinct nature produces poles in $\e$ \cite{Smirnov:2004ym}. 
The asterisk in the Euler $\Gamma$ function means that the most left of the right poles changes its nature 
and does not contribute to the right residues any more. The contour $C$ includes the vertical line 
that is a bit on the left of the imaginary axis in both the integrals on r.h.s. of Eq.(\ref{res}). 
Here it is necessary to take into account that in the Cauchy theorem the contour orientation is counterclockwise, however in the 
MB transformation of the integral (\ref{J-arb}) we have clockwise orientation of the contour. The term  $R$ on the 
r.h.s. of Eq.(\ref{res}) is investigated in Appendix A.

After making simple algebra in integrand, integral $I_1$ can be written as 
\begin{eqnarray}
I_1 = \oint_C dz_2~dz_3~ \frac{x^{z_2}~y^{z_3}}{[23]^2} \left\{ \G^2 \le -z_3 \ri \G \le -z_2 \ri  
\G^* \le -z_2 -1 \ri \G^2 \le z_3 + z_2 + 1 \ri \le z_2 + z_3  + 1\ri \right\}  \no\\
= - \oint_C dz_2~dz_3~ \frac{x^{z_2}~y^{z_3}}{[23]^2} \left\{ \G^2 \le -z_3 \ri \G^2 \le -z_2 \ri\G^2 \le z_3 + z_2 + 1 \ri \right\} 
- I_3, \no\\
I_3 \equiv \oint_C dz_2~dz_3~ \frac{x^{z_2}~y^{z_3}}{[23]^2} \left\{ \G \le -z_3 \ri \G \le -z_3 +1\ri  \G \le -z_2 \ri  
\G^* \le -z_2 -1 \ri \G^2 \le z_3 + z_2 + 1 \ri \right\} \label{I3} 
\end{eqnarray}
For such a contour $C$ the identity $\G^* \le -z_2 -1 \ri \le z_2 + 1 \ri = - \G \le -z_2 \ri$ is valid. This identity 
allows to express one of the integrals that appear on the r.h.s.  of Eq.(\ref{I3}) in terms of $J(1,1,1).$ The integral $I_3$ 
is investigated in Appendix B and the result for $I_1$ is 
\begin{eqnarray*}
I_1 = - \frac{[23] + [31]- [12]}{2[23][31]} J(1,1,1)  - \frac{(\ln x -2\psi(1))(1-y)}{2[23][31]} I_4 - \frac{1-y}{[23][31]} I_5, 
\end{eqnarray*}
where one-fold MB integrals $I_4$ and $I_5$ are defined in Appendix B. Thus, Eq.(\ref{res}) can be written as 
\begin{eqnarray*}
J(\L,\L,2\L) = \frac{1}{(2\e-1)}\frac{1}{\G(1-\e)}\frac{[12]^{-\e}}{[31]^{1-2\e}[23]^{1-2\e}}
\Bigl[-\frac{1}{\e}\frac{\G(1+\e)\G(1-2\e)}{\G(1-\e)} \Bigr.\\ 
+ \Bigl.\e\left[-2 I_2  + (1-y)\le\ln x -2\psi(1)\ri I_4 + 2(1-y) I_5  +  \le [23] + [31]- [12]\ri J(1,1,1) \right]\Bigr] + o(\e). 
\end{eqnarray*}
With help of Eq.(12) of Ref.\cite{Cvetic:2006iu} 
\begin{eqnarray*}
\Gamma(1+a\e)=\exp \left[-\gamma a\e +\sum_{k=2}^{\infty}\frac{\zeta(k)}{k}(-a\e)^k \right] = 
\exp \left[-\gamma a\e + \frac{\zeta(2)}{2}(a\e)^2 \right] + o(\e^2),
\end{eqnarray*}
we obtain
\begin{eqnarray*}
\frac{\G(1+\e)\G(1-2\e)}{\G(1-\e)} = \exp\left[2\zeta(2)\e^2 \right] + o(\e^2) = 1 + 2\zeta(2)\e^2 + o(\e^2).
\end{eqnarray*}
Thus, we have derived 
\begin{eqnarray*}
J(\L,\L,2\L) = \frac{1}{(2\e-1)}\frac{1}{\G(1-\e)}\frac{[12]^{-\e}}{[31]^{1-2\e}[23]^{1-2\e}} \\\
\times \Bigl[-\frac{1}{\e} + \e\left[-2\zeta(2) - 2 I_2  + (1-y)\le\ln x -2\psi(1)\ri I_4 + 2(1-y) I_5  + \right.\Bigr.\\
\Bigl. \left. \le [23] + [31]- [12]\ri  J(1,1,1) \right]\Bigr] + o(\e) \\
= \frac{1}{(2\e-1)}\frac{1}{\G(1-\e)}\frac{[12]^{-\e}}{[31]^{1-2\e}[23]^{1-2\e}} \\ 
\times \left[-\frac{1}{\e} + \e\left[\ln y \ln \frac{y}{x}  + \le [23] + [31]- [12]\ri J(1,1,1) \right]\right] + o(\e),
\end{eqnarray*}
or in other words the identity is valid
\begin{eqnarray*}
-2\zeta(2) - 2 I_2  + (1-y)\le\ln x -2\psi(1)\ri I_4 + 2(1-y) I_5 = \ln y \ln \frac{y}{x} 
\end{eqnarray*}

This identity can be verified in a straightforward way by calculating the residues. One-fold MB integrals cannot produce any UD function 
and are represented in terms of logarithms.

\section{Mellin-Barnes transform for $J(\L-\delta,\L,2\L+\delta)$}

The MB transform (\ref{J-arb}) of the integral in Eq.(\ref{F}) is written as    
\begin{eqnarray}
J(\L - \delta,\L,2\L+\delta) \no\\ 
= \frac{2\e}{\G(\L)\G(\L-\delta)\G(2\L+\delta)\G(1+2\e)} 
\oint_C dz_2~dz_3~ \frac{x^{z_2}~y^{z_3}}{[23]^{2-3\e}} \left\{\G \le -z_3 \ri \G \le -z_2 \ri \right.\no\\
\left.\G \le -z_2 -1 +2\e\ri \G \le -z_3 + \e + \delta \ri 
\G \le z_3 + z_2 + 2 -3\e \ri \G \le z_2 + z_3  + 1 - \e - \delta\ri \right\}.\label{JM-part}
\end{eqnarray}
We repeat all the steps made in the previous section for  the integral $J(\L,\L,2\L)$ in order to extract MB transform of
$J(1,1,1).$ The main point is that all the operations with the contour of integration in $z_2$ and $z_3$ complex planes
are completely the same as for the  integral $J(\L,\L,2\L).$  In case of Eq.(\ref{JM-part}) the most left of the right poles 
is the pole $z_2 = -1 + 2\e$ in $z_2$ plane and the rest of the right poles are again situated on the right half-planes of 
$z_2$ and $z_3$ complex planes. Separating this pole from the set of the right residues in $z_2$ plane, we have 
\begin{eqnarray}
J(\L - \delta,\L,2\L+\delta) = R' +  \frac{2\e}{\G(1-\delta)\G(2 +\delta)}  I'_1 + o(\e), \label{resM}\\
R' \equiv   \frac{2\e\G(1-2\e)}{\G(\L)\G(\L-\delta)\G(2\L+\delta)\G(1+2\e)} \frac{x^{-1+2\e}}{[23]^{2-3\e}} \no\\
\times \oint_C dz_3~ y^{z_3}\left\{\G \le -z_3 \ri \G \le -z_3 + \e + \delta \ri 
\G \le z_3 + 1 - \e \ri \G \le z_3  + \e - \delta \ri \right\},  \no\\
I'_1 \equiv   \oint_C dz_2~dz_3~ \frac{x^{z_2}~y^{z_3}}{[23]^2} \left\{ \G \le -z_3 \ri \G \le -z_3 +\delta \ri 
\G \le -z_2 \ri  \G^* \le -z_2 -1 \ri  \right. \no\\ 
\left.\G \le z_3 + z_2 + 2 \ri  \G \le z_2 + z_3  + 1 - \delta \ri \right\}    \no
\end{eqnarray}
The contour of integration $C$ in Eq.(\ref{resM}) separates left and right poles in $z_2$ and $z_3$ complex planes 
and in $I'_1$ coincides with the contour $C$ in Eq.(\ref{res}). In integral $I'_1$ 
we have removed the dependence on $\e$ in the integrand since no gluing effect between two distinct poles of 
different nature appears, and we have taken into account  a clockwise orientation of the contour. The term  $R'$ on the 
r.h.s. of Eq.(\ref{resM}) is investigated in Appendix C.

After making simple algebra in integrand, integral $I'_1$ can be written as 
\begin{eqnarray}
I'_1 = \oint_C dz_2~dz_3~ \frac{x^{z_2}~y^{z_3}}{[23]^2} \left\{ \G \le -z_3 \ri \G \le -z_3 +\delta \ri  \G \le -z_2 \ri  
\G^* \le -z_2 -1 \ri  \right. \no\\
\left.  \G \le z_3 + z_2 + 1 \ri \le z_2 + z_3  + 1\ri \G \le z_2 + z_3  + 1 - \delta \ri \right\}  \no\\
= - \oint_C dz_2~dz_3~ \frac{x^{z_2}~y^{z_3}}{[23]^2} \left\{ \G \le -z_3 \ri  \G \le -z_3 +\delta \ri \right.\no\\
\left. \G^2 \le -z_2 \ri\G \le z_3 + z_2 + 1 \ri \G \le z_2 + z_3  + 1 - \delta \ri\right\}  - I'_3 \no\\
=  -\frac{\G(1 - \delta)\G(1 + \delta)}{[23]^{1+\delta}}J(1,1-\delta,1) - I'_3, \label{I3M}\\
I'_3 \equiv \oint_C dz_2~dz_3~ \frac{x^{z_2}~y^{z_3}}{[23]^2} \left\{ \G \le -z_3 +\delta\ri \G \le -z_3 +1\ri  
\G \le -z_2 \ri \right.\no\\ 
\left. \G^* \le -z_2 -1 \ri \G \le z_3 + z_2 + 1 \ri \G \le z_2 + z_3  + 1 - \delta \ri \right\}  \no
 \end{eqnarray}
For such a contour $C$ the identity $\G^* \le -z_2 -1 \ri \le z_2 + 1 \ri = - \G \le -z_2 \ri$ is valid. This identity 
allows to express one of the integrals that appear in Eq.(\ref{I3M}) in terms of $J(1,1-\delta,1).$
Integral $I'_3$ can be represented in terms of $J(1,1-\delta,1)$ too. Here, it is necessary to mention that the derivative of 
$J(1,1-\delta,1)$ with respect to $\delta$ is reduced to the second UD function \cite{Usyukina:1992jd,Usyukina:1993ch}. 
The integral $I'_3$ is investigated in Appendix D and the result for the $I_1'$ is
\begin{eqnarray*}
I'_1 = - \frac{[23]+[31]-[12]}{2[31][23]^{1+\delta}} \G(1 - \delta)\G(1 + \delta)J(1,1-\delta,1) 
- \frac{(\ln x -2\psi(1))(1-y)}{2[23][31]} I'_4 \\
- \frac{1-y}{2[23][31]} I'_5 + \frac{\delta}{[23]^2}I'_6 + o(\delta),  
\end{eqnarray*}
where one-fold MB integrals $I'_4$ and $I'_5$ are defined in Appendix D. Thus, we obtain
\begin{eqnarray}
J(\L-\delta,\L,2\L+\delta) = \frac{1}{\e-\delta}\frac{\G(1+\e-\delta)\G(1-2\e)}{(1+\delta-2\e)\G(1-\e)\G(1-\e-\delta)} 
\frac{[12]^{\delta-\e}}{[31]^{1-2\e}[23]^{1+\delta-2\e}} \no\\
- \e\frac{[23]+[31]-[12]}{[31][23]^{1+\delta}} J(1,1-\delta,1) + \frac{2\e\delta}{[23]^2}I'_6 \no\\
- \frac{2\e}{\G(1-\delta)\G(2 +\delta)}\left[ -\frac{1}{[31][23]}I'_2 + \frac{(\ln x -2\psi(1))(1-y)}{2[23][31]} I'_4 
+ \frac{1-y}{2[23][31]} I'_5  \right]  + o(\e)+ o(\delta). \label{final}
\end{eqnarray}
The definitions for the integrals are collected below, 
\begin{eqnarray*}
I'_2 = \oint_{C} dz_3~ y^{z_3} \left\{ \G\le -z_3 \ri \G(-z_3+\delta) \G \le z_3 + 1 \ri \G^* \le z_3 - \delta\ri \right\}, \\
I'_4 = \oint_C~dz_3~y^{z_3} \G(-z_3) \G(-z_3+\delta) \G(1+z_3) \G(1+z_3-\delta), \\
I'_5 =  \oint_C~dz_3~y^{z_3} \G(-z_3) \G(-z_3+\delta) \G(1+z_3) \G(1+z_3-\delta) \le \psi(1+z_3) + \psi(1+z_3-\delta) \ri,\\
I'_6 = \oint_C dz_2~dz_3~ x^{z_2}~y^{z_3} \G^2 \le -z_3 \ri  {\G^*} \le -z_2 -1 \ri \G \le -z_2 \ri  \G^2 \le z_3 + z_2 + 1 \ri
\end{eqnarray*}
The integral $I'_6$ is a two-fold MB integral. It can produce the UD functions, and it can be expressed in terms of them. 
The key observation that such a presentation in terms of the UD functions exists is that the derivative of $I'_6$ 
with respect to $y$ or $x$  produces $J(1,1,1).$  From the rules of Refs.\cite{Cvetic:2007fp,Cvetic:2007ds} we know that 
this is a property of the first UD function.  Due to this, it is expected that $I'_6$ can be expressed 
in terms of the UD functions. The one-fold integrals that appear in the final expression cannot produce the UD functions, and  
their combination in the result  (\ref{final}) is reduced to the logarithmic function. Thus, we have shown via MB transform 
that the integrals with logarithms of the type (\ref{obj-mod}) should be reduced to the UD functions, logarithmic functions and powers
of space-time intervals. For this particular problem, the advantage of the method of MB transform  is that we even 
did not integrate anything or sum any series. The result is obtained by transforming the contour of integration in the complex plane.

\section{Conclusion}

In this paper we have argued via MB transform  that three point scalar integrals in the position space which contribute 
to the Green functions in ${\cal N} = 4$ supersymmetric Yang-Mills theory  and in other four-dimensional massless 
theories can be represented in terms of UD functions corresponding to the scalar ladder three point diagrams in four 
spacetime dimensions.  It is an important observation since  they possess remarkable properties  with respect to various  
integral transformations  and  can serve as blocks  in terms of which ${\cal N} = 4$  supersymmetric theory  
is built \cite{Drummond:2006rz,Kondrashuk:2008xq,Kondrashuk:2008ec}. One of these properties is that the UD 
functions transform to themselves under the Fourier transform \cite{Kondrashuk:2008xq,Kondrashuk:2009us,Kondrashuk:2008ec}.  
The most simple proof of this fact can be found in  Appendix E.

In Ref. \cite{Cvetic:2007fp} it was mentioned that this property can serve as an indication  that the conformal symmetries 
in the position space and in the momentum space  can be unified.  In Ref. \cite{Hodges:2009hk} the unification 
was found by means of extension of the twistor language used for amplitudes \cite{Bern:2005iz,Bern:2006ew} 
to a more general form in which the momentum  component and position component of twistors participate on equal foot.

Taking into account the symmetry with respect to Fourier transform, we realize importance of making reduction 
of the correlators to the UD functions in the momentum space for ${\cal N} = 4$  supersymmetric Yang-Mills 
theory as it has been done in Ref. \cite{Drummond:2006rz}. For that theory by using unitarity methods 
it has been demonstrated that only the diagrams created due to the ``rung rule'' contribute to the four-point gluon amplitude 
up to three-loop level \cite{Bern:2005iz}.  These ``rung rule'' diagrams in the momentum space are represented by the same 
UD function Refs.~\cite{Usyukina:1992jd,Usyukina:1993ch} of conformally invariant ratios of squares of 
external momenta and their linear combinations.  This allowed to classify all conformally invariant contributions 
in the momentum space up to four-loop level \cite{Drummond:2006rz,Bern:2006ew,Nguyen:2007ya}. 
The same conformal symmetry in the momentum space appears in Alday-Maldacena approach \cite{Alday:2007hr} to calculate gluon
scattering amplitude on the string side at strong coupling. Conformal invariant contributions to four-point gluon 
amplitude  reproduce the known results for the anomalous dimensions of twist-two operators in maximally supersymmetric 
Yang-Mills theory \cite{Kotikov:2002ab,Kotikov:2003fb,Kotikov:2004er}.

\section{Acknowledgments}

I. Kondrashuk was supported by Fondecyt (Chile) grants  1040368, 1050512 and by DIUBB grant (UBB, Chile) 082609. E.A. Notte-Cuello
was supported by Direcci\'on de Investigaci\'on de la Universidad de La Serena, DIULS CD091501. 
This paper is based on I.K.'s talks at El Congreso de Matem\'atica Capricornio, COMCA 2009, Antofagasta, Chile 
and at DMFA seminar, UCSC, Concepci\'on, Chile, and he thanks the organizers for opportunity to present these results.

\begin{appendix}

\section[]{}
\setcounter{equation}{0}
\label{App:A}

In this Appendix  we treat $R$ term in Eq.(\ref{res}) because it can produce poles in $\e.$ Indeed, there is a gluing effect 
between the left pole at the point $z_3 = - \e$ produced by $\G(z_3+\e)$ and the right pole produced by $\G(-z_3)$ at 
the point $z_3 = 0$  in the limit $\e \rar 0.$  It makes the operations of contour integration and decomposition of the 
integrand in powers of $\e$ non-commutative. To make these operations commutative, we need to convert the left pole at the point 
$z_3 = - \e$ into the right pole by changing the contour of integration to include it in the set of right poles in $z_3$ plane. 
Thus, we have to write 
\begin{eqnarray}
\oint_C dz_3~ y^{z_3} \left\{\G \le -z_3 \ri \G \le -z_3 + \e \ri \G \le z_3 + 1 - \e \ri \G \le z_3  + \e \ri \right\}  \no\\
= \oint_{C'} dz_3~ y^{z_3} \left\{\G \le -z_3 \ri \G \le -z_3 + \e \ri \G \le z_3 + 1 - \e \ri \G^* \le z_3  + \e \ri \right\}  \no\\
+ y^{-\e}\G(\e)\G(2\e)\G(1-2\e) \no\\
= \oint_{C} dz_3~ y^{z_3} \left\{\G^2 \le -z_3 \ri \G \le z_3 + 1 \ri \G^* \le z_3 \ri \right\} 
+ y^{-\e}\G(\e)\G(2\e)\G(1-2\e) + o(\e)  \no\\
\equiv I_2 + y^{-\e}\G(\e)\G(2\e)\G(1-2\e) + o(\e). \label{J-arb-2}
\end{eqnarray}
The difference between the contour $C$ and $C'$ is that $C'$ includes the pole at the point $z_3 = - \e.$ 
The asterisk in the first term on the r.h.s. 
means that this most right of the set of the left poles changes its nature and contributes to 
the right residues.  In this term we 
take the limit $\e \rar 0,$ this converts $C'$ back to $C.$  This limit is a well-defined operation after the contour modification done
since no effect of gluing left and right poles appear in the first term on the r.h.s. of Eq.(\ref{J-arb-2}). 
The second term on the r.h.s. is the contribution of the residue at the point $z_3 = - \e.$  It is taken with the positive 
sign due to clockwise orientation of contour $C,$ 
\begin{eqnarray}
R = 
\frac{2\e}{(1-2\e)\G^2(1-\e)\G(1+2\e)} \frac{x^{-1+2\e}}{[23]^{2-3\e}}\left[I_2  + y^{-\e}\G(\e)\G(2\e)\G(1-2\e) + o(\e) \right] \no\\
= \frac{2\e}{[31][23]}I_2 + \frac{2\e\G(\e)\G(2\e)\G(1-2\e)}{(1-2\e)\G^2(1-\e)\G(1+2\e)} \frac{x^{-1+2\e}y^{-\e}}{[23]^{2-3\e}}  
+ o(\e)  \no\\
= -\frac{1}{\e}\frac{\G(1+\e)\G(1-2\e)}{(2\e-1)\G^2(1-\e)} \frac{[12]^{-\e}}{[31]^{1-2\e}[23]^{1-2\e}} + \frac{2\e}{[31][23]}I_2  
+ o(\e). 
\label{FT}
\end{eqnarray}

\section[]{}
\setcounter{equation}{0}
\label{App:B}

After a little algebra, the integral $I_3$ can be represented in terms of $J(1,1,1)$ too,
\begin{eqnarray*}
I_3 = \oint_C dz_2~dz_3~ \frac{x^{z_2}~y^{z_3}}{[23]^2} \left\{ \G \le -z_3 \ri \G \le -z_3 +1\ri  \G \le -z_2 \ri  
\G^* \le -z_2 -1 \ri \G^2 \le z_3 + z_2 + 1 \ri \right\} \\
= \frac{1}{2}\left[2 I_3 + \frac{1}{[23]}J(1,1,1) - \frac{1}{[23]}J(1,1,1)\right]  \\
= \frac{1}{2}\left[\oint_C dz_2~dz_3~ \frac{x^{z_2}~y^{z_3}}{[23]^2} \left\{ \G^2 \le -z_3 \ri 
{\G^*}^2 \le -z_2 -1 \ri \le 2z_2 + 2 \ri z_3 ~\G^2 \le z_3 + z_2 + 1 \ri \right\} \right.\\ 
+ \left.\oint_C dz_2~dz_3~ \frac{x^{z_2}~y^{z_3}}{[23]^2} \left\{ \G^2 \le -z_3 \ri {\G^*}^2 \le -z_2 -1 \ri \le z_2 + 1\ri^2 
\G^2 \le z_3 + z_2 + 1 \ri \right\} -  \frac{1}{[23]}J(1,1,1) \right]  \\
= \frac{1}{2}\left[\oint_C dz_2~dz_3~ \frac{x^{z_2}~y^{z_3}}{[23]^2} \left\{ \G^2 \le -z_3 \ri 
{\G^*}^2 \le -z_2 -1 \ri \le 2(z_2 + 1)z_3 + (z_2+1)^2 \ri ~\G^2 \le z_3 + z_2 + 1 \ri \right\} \right.\\ 
\left. -  \frac{1}{[23]}J(1,1,1) \right] \\
= \frac{1}{2}\left[\oint_C dz_2~dz_3~ \frac{x^{z_2}~y^{z_3}}{[23]^2} \left\{ \G^2 \le -z_3 \ri 
{\G^*}^2 \le -z_2 -1 \ri \le (z_3+ z_2+1)^2 - z_3^2 \ri ~\G^2 \le z_3 + z_2 + 1 \ri \right\} \right.\\ 
\left. -  \frac{1}{[23]}J(1,1,1) \right] \\
= \frac{1}{2}\left[\oint_C dz_2~dz_3~ \frac{x^{z_2}~y^{z_3}}{[23]^2} \left\{ \G^2 \le -z_3 \ri 
{\G^*}^2 \le -z_2 -1 \ri \G^2 \le z_3 + z_2 + 2 \ri \right\} \right.\\ 
- \left. \oint_C dz_2~dz_3~ \frac{x^{z_2}~y^{z_3}}{[23]^2} \left\{ \G^2 \le -z_3 + 1 \ri 
{\G^*}^2 \le -z_2 -1 \ri \G^2 \le z_3 + z_2 + 1 \ri \right\} -  \frac{1}{[23]}J(1,1,1) \right]  \\
= \frac{1}{2}\left[\oint_C dz_2~dz_3~ \frac{x^{z_2-1}~y^{z_3}}{[23]^2} \left\{ \G^2 \le -z_3 \ri 
{\G^*}^2 \le -z_2 \ri \G^2 \le z_3 + z_2 + 1 \ri \right\} \right.\\ 
- \left. \oint_C dz_2~dz_3~ \frac{x^{z_2-1}~y^{z_3+1}}{[23]^2} \left\{ \G^2 \le -z_3 \ri 
{\G^*}^2 \le -z_2 \ri \G^2 \le z_3 + z_2 + 1 \ri \right\} -  \frac{1}{[23]}J(1,1,1) \right] \\
= \frac{1}{2[23]} \left[\le \frac{1}{x} - \frac{y}{x} \ri 
\oint_C dz_2~dz_3~ \frac{x^{z_2}~y^{z_3}}{[23]} \left\{ \G^2 \le -z_3 \ri 
{\G^*}^2 \le -z_2 \ri \G^2 \le z_3 + z_2 + 1 \ri \right\} -  J(1,1,1) \right]  \\
= \frac{1}{2[23]} \left[\le\frac{1}{x} - \frac{y}{x} \ri \Bigl( J(1,1,1)  \Bigr.\right.\\
+ \Bigl.\left. \frac{1}{[23]}\oint_C~dz_3~y^{z_3}\G^2(-z_3)\G^2(1+z_3) \le -2\psi(1) + 2\psi(1+z_3) + \ln x\ri \Bigl) 
-  J(1,1,1) \right]  \\
= \frac{1}{2[23]} \left[\frac{1-x-y}{x} J(1,1,1) \right. \\
+ \left. \frac{1-y}{x[23]}\oint_C~dz_3~y^{z_3}\G^2(-z_3)\G^2(1+z_3) \le -2\psi(1) + 2\psi(1+z_3) + \ln x\ri  \right]  \\
= \frac{[23]-[31]-[12]}{2[23][31]} J(1,1,1) + \frac{(\ln x -2\psi(1))(1-y)}{2[23][31]}\oint_C~dz_3~y^{z_3}\G^2(-z_3)\G^2(1+z_3)  \\
+ \frac{1-y}{[23][31]}\oint_C~dz_3~y^{z_3}\G^2(-z_3)\G^2(1+z_3) \psi(1+z_3)   \\
\equiv \frac{[23]-[31]-[12]}{2[23][31]} J(1,1,1) + \frac{(\ln x -2\psi(1))(1-y)}{2[23][31]} I_4 + \frac{1-y}{[23][31]} I_5.  
\end{eqnarray*}
Integrals $I_4$ and $I_5$ are one-fold MB integrals in $z_3$ plane produced due to to inclusion the residue at  
$z_2 = 0$ in $z_2$ plane in order to recover $J(1,1,1)$ construction. Again, we have added the residue instead of removing 
it due to clockwise orientation of $C,$ where the definitions of the one-fold MB integrals can be collected as   
\begin{eqnarray*}
I_2 = \oint_{C} dz_3~ y^{z_3} \left\{\G^2 \le -z_3 \ri \G \le z_3 + 1 \ri \G^* \le z_3 \ri \right\} \\
I_4 = \oint_C~dz_3~y^{z_3}\G^2(-z_3)\G^2(1+z_3) \\
I_5 =  \oint_C~dz_3~y^{z_3}\G^2(-z_3)\G^2(1+z_3) \psi(1+z_3).
\end{eqnarray*}

\section[]{}
\setcounter{equation}{0}
\label{App:C}

In this Appendix we analyse the $R'$ term in Eq.(\ref{resM}) because it can produce poles in $\e.$ Again, we obtain complete 
analogy with the case of $J(\L,\L,2\L)$ in which the first term in Eq.(\ref{res}) generates pole in $\e.$ 
It is supposed that $\delta$ is a small real positive variable, $\delta > {\rm Re~\e}.$  We observe the gluing effect 
between the left pole at the point $z_3 = \delta - \e$ produced by $\G(z_3+\e - \delta)$ and the right pole produce by 
$\G(-z_3 + \delta +\e)$ at the point $z_3 = \delta + \e$  in the limit $\e \rar 0.$ 
In other words, the operations of contour integration and decomposition of the integrand in powers of $\e$ do not commute 
for the first term in Eq.(\ref{resM}).   To make these operations commutative, we need to convert the left pole at the point 
$z_3 = \delta - \e$ into the right pole by changing the contour of integration to include it in the set of right poles in 
$z_3$ plane. Thus, we obtain
\begin{eqnarray}
\oint_C dz_3~ y^{z_3} \left\{\G \le -z_3 \ri  \G \le -z_3 + \e + \delta \ri  \G \le z_3 + 1 - \e \ri 
\G \le z_3  + \e - \delta \ri  \right\}  \no\\
= \oint_{C'} dz_3~ y^{z_3} \left\{\G \le -z_3 \ri   \G \le -z_3 + \e + \delta \ri   \G \le z_3 + 1 - \e \ri 
\G^* \le z_3  + \e -\delta \ri \right\}  \no\\
+~ y^{\delta -\e}~\G(\e-\delta)\G(2\e)\G(1+\delta-2\e) \no \\
= \oint_{C} dz_3~ y^{z_3} \left\{\G\le -z_3 \ri \G\le -z_3 +\delta \ri \G \le z_3 + 1 \ri \G^* \le z_3 - \delta \ri \right\} \no\\ 
+~ y^{\delta -\e}\G(\e-\delta)\G(2\e)\G(1+\delta-2\e) + o(\e) \no \\
\equiv I'_2 +  y^{\delta -\e}\G(\e-\delta)\G(2\e)\G(1+\delta-2\e) + o(\e). \label{J-arb-2M}
\end{eqnarray}
The difference between the contour $C$ and $C'$ is that $C'$ includes the pole at the point $z_3 = \delta - \e.$ 
The asterisk in the second line means that this most right of the set of the left poles changes its nature and contributes to 
the right residues. The contour $C'$ of  Eq.(\ref{J-arb-2M}) is the vertical line and it coincides with the contour $C$ of 
Eq.(\ref{J-arb-2}).  In integral $I'_2$ we take the limit $\e \rar 0.$   This limit is a well-defined operation after the 
contour modification done since no effect of gluing left and right poles appear. The second term on the r.h.s. of Eq.(\ref{J-arb-2M}) 
is the contribution of the residue at the point $z_3 = \delta - \e.$  It is taken with the positive sign due to  
clockwise orientation of contour $C.$ The first term  of Eq.(\ref{resM})  is transformed to 
\begin{eqnarray}
R' = \frac{2\e\G(1-2\e)}{\G(\L)\G(\L-\delta)\G(2\L+\delta)\G(1+2\e)} \frac{x^{-1+2\e}}{[23]^{2-3\e}} \no\\
\times \left[I'_2 +  y^{\delta -\e}\G(\e-\delta)\G(2\e)\G(1+\delta-2\e) + o(\e) \right]  \no\\
= \frac{1}{\G(1-\delta)\G(2+\delta)} \frac{2\e}{[31][23]}I'_2 + 
\frac{2\e\G(\e-\delta)\G(2\e)\G(1+\delta-2\e)\G(1-2\e)}{\G(\L)\G(\L-\delta)\G(2\L+\delta)\G(1+2\e)} 
\frac{x^{-1+2\e}y^{\delta-\e}}{[23]^{2-3\e}} + o(\e)  \no\\
= \frac{1}{\e-\delta}\frac{\G(1+\e-\delta)\G(1-2\e)}{(1+\delta-2\e)\G(1-\e)\G(1-\e-\delta)} 
\frac{[12]^{\delta-\e}}{[31]^{1-2\e}[23]^{1+\delta-2\e}} \no\\
+ \frac{1}{\G(1-\delta)\G(2+\delta)}\frac{2\e}{[31][23]}I'_2 + o(\e).\label{FTM} 
\end{eqnarray}
In the case  $\delta = 0$ we reproduce Eq.(\ref{FT}).

\section[]{}
\setcounter{equation}{0}
\label{App:D}

\begin{eqnarray*}
I'_3 = \oint_C dz_2~dz_3~ \frac{x^{z_2}~y^{z_3}}{[23]^2} \left\{ \G \le -z_3 +\delta\ri \G \le -z_3 +1\ri  \G \le -z_2 \ri 
\G^* \le -z_2 -1 \ri\right.\\ 
\left. \G \le z_3 + z_2 + 1  \ri\G \le z_3 + z_2 + 1 - \delta \ri \right\}  \\
= \frac{1}{2}\left[2 I'_3 +  \frac{\G(1 - \delta)\G(1 + \delta)}{[23]^{1+\delta}}J(1,1-\delta,1) 
-  \frac{\G(1 - \delta)\G(1 + \delta)}{[23]^{1+\delta}}J(1,1-\delta,1) \right]  \\
= \frac{1}{2}\left[\oint_C dz_2~dz_3~ \frac{x^{z_2}~y^{z_3}}{[23]^2} \left\{ \G \le -z_3 \ri \G \le -z_3 +\delta\ri
{\G^*}^2 \le -z_2 -1 \ri \le 2z_2 + 2 \ri z_3 \right.\right.\\ 
\left.\left. ~\G \le z_3 + z_2 + 1  \ri\G \le z_3 + z_2 + 1 - \delta \ri\right\} \right.\\ 
+ \left.\oint_C dz_2~dz_3~ \frac{x^{z_2}~y^{z_3}}{[23]^2} \left\{ \G \le -z_3 \ri \G \le -z_3 + \delta\ri 
{\G^*}^2 \le -z_2 -1 \ri \right.\right.\\ 
\left. \left.\le z_2 + 1\ri^2 \G \le z_3 + z_2 + 1  \ri\G \le z_3 + z_2 + 1 - \delta \ri\right\} \right.\\
- \left.  \frac{\G(1 - \delta)\G(1 + \delta)}{[23]^{1+\delta}}J(1,1-\delta,1)  \right]  \\
= \frac{1}{2}\left[\oint_C dz_2~dz_3~ \frac{x^{z_2}~y^{z_3}}{[23]^2} \left\{ \G \le -z_3 \ri \G \le -z_3 + \delta\ri 
{\G^*}^2 \le -z_2 -1 \ri  \right.\right.\\ 
\left.\left. \le 2(z_2 + 1)z_3 + (z_2+1)^2 \ri ~\G \le z_3 + z_2 + 1 \ri \G \le z_3 + z_2 + 1 -\delta \ri \right\} \right.\\ 
\left. -  \frac{\G(1 - \delta)\G(1 + \delta)}{[23]^{1+\delta}}J(1,1-\delta,1)\right]  \\
= \frac{1}{2}\left[\oint_C dz_2~dz_3~ \frac{x^{z_2}~y^{z_3}}{[23]^2} \left\{ \G \le -z_3 \ri \G \le -z_3 + \delta\ri 
{\G^*}^2 \le -z_2 -1 \ri  \right.\right.\\ 
\left.\left. \le (1+z_2+z_3)^2 - z_3^2  - \delta(z_2+1)  + \delta(z_2+1)\ri 
~\G \le z_3 + z_2 + 1 \ri \G \le z_3 + z_2 + 1 -\delta \ri \right\} \right.\\ 
\left. -  \frac{\G(1 - \delta)\G(1 + \delta)}{[23]^{1+\delta}}J(1,1-\delta,1)\right]  \\
= \frac{1}{2}\left[\oint_C dz_2~dz_3~ \frac{x^{z_2}~y^{z_3}}{[23]^2} \left\{ \G \le -z_3 \ri \G \le -z_3 + \delta\ri 
{\G^*}^2 \le -z_2 -1 \ri  \right.\right.\\ 
\left.\left. \le (1+z_2+z_3)(1+z_2+z_3-\delta) - z_3(z_3-\delta)  + \delta(z_2+1)\ri 
~\G \le z_3 + z_2 + 1 \ri \G \le z_3 + z_2 + 1 -\delta \ri \right\} \right.\\ 
\left. -  \frac{\G(1 - \delta)\G(1 + \delta)}{[23]^{1+\delta}}J(1,1-\delta,1)\right]  \\
= \frac{1}{2}\left[\oint_C dz_2~dz_3~ \frac{x^{z_2}~y^{z_3}}{[23]^2} \left\{ \G \le -z_3 \ri \G \le -z_3 + \delta\ri 
{\G^*}^2 \le -z_2 -1 \ri  \right.\right.\\ 
\left.\left. \G \le z_3 + z_2 + 2 \ri \G \le z_3 + z_2 + 2 -\delta \ri \right\} \right.\\
- \left.\oint_C dz_2~dz_3~ \frac{x^{z_2}~y^{z_3}}{[23]^2} \left\{ \G \le -z_3 +1\ri \G \le -z_3 + 1+\delta\ri 
{\G^*}^2 \le -z_2 -1 \ri  \right.\right.\\ 
\left.\left. ~\G \le z_3 + z_2 + 1 \ri \G \le z_3 + z_2 + 1 -\delta \ri \right\} \right. \\
- \left. \delta \oint_C dz_2~dz_3~ \frac{x^{z_2}~y^{z_3}}{[23]^2} \left\{ \G^2 \le -z_3 \ri  
{\G^*} \le -z_2 -1 \ri  \G \le -z_2 \ri  \G^2 \le z_3 + z_2 + 1 \ri \right\} \right.\\
\left. -  \frac{\G(1 - \delta)\G(1 + \delta)}{[23]^{1+\delta}}J(1,1-\delta,1) + o(\delta)\right]  \\
= \frac{1}{2}\left[\oint_C dz_2~dz_3~ \frac{x^{z_2-1}~y^{z_3}}{[23]^2} \left\{ \G \le -z_3 \ri \G \le -z_3 + \delta\ri 
{\G^*}^2 \le -z_2 \ri  \right.\right.\\ 
\left.\left. \G \le z_3 + z_2 + 1 \ri \G \le z_3 + z_2 + 1 -\delta \ri \right\} \right.\\
- \left.\oint_C dz_2~dz_3~ \frac{x^{z_2-1}~y^{z_3+1}}{[23]^2} \left\{ \G \le -z_3\ri \G \le -z_3 +\delta\ri 
{\G^*}^2 \le -z_2  \ri  \right.\right.\\ 
\left.\left. ~\G \le z_3 + z_2 + 1 \ri \G \le z_3 + z_2 + 1 -\delta \ri \right\} \right. \\
- \left. \delta \oint_C dz_2~dz_3~ \frac{x^{z_2}~y^{z_3}}{[23]^2} \left\{ \G^2 \le -z_3 \ri  
{\G^*} \le -z_2 -1 \ri  \G \le -z_2 \ri  \G^2 \le z_3 + z_2 + 1 \ri \right\} \right.\\
\left. -  \frac{\G(1 - \delta)\G(1 + \delta)}{[23]^{1+\delta}}J(1,1-\delta,1) + o(\delta)\right]  \\
= \frac{1}{2} \left[\le \frac{1}{x} - \frac{y}{x} \ri 
\oint_C dz_2~dz_3~ \frac{x^{z_2}~y^{z_3}}{[23]^2} \left\{ \G \le -z_3 \ri \G \le -z_3 + \delta\ri 
{\G^*}^2 \le -z_2 \ri  \right.\right.\\ 
\left.\left. \G \le z_3 + z_2 + 1 \ri \G \le z_3 + z_2 + 1 -\delta \ri \right\} \right.\\
- \left. \delta \oint_C dz_2~dz_3~ \frac{x^{z_2}~y^{z_3}}{[23]^2} \left\{ \G^2 \le -z_3 \ri  
{\G^*} \le -z_2 -1 \ri  \G \le -z_2 \ri  \G^2 \le z_3 + z_2 + 1 \ri \right\} \right.\\
\left. -  \frac{\G(1 - \delta)\G(1 + \delta)}{[23]^{1+\delta}}J(1,1-\delta,1) + o(\delta)\right]  \\
= \frac{1}{2} \left[\frac{1-x-y}{x}  \frac{\G(1 - \delta)\G(1 + \delta)}{[23]^{1+\delta}}J(1,1-\delta,1) \right. \\
+ \left.\frac{1}{[23]^2}\frac{1-y}{x} \oint_C~dz_3~y^{z_3}\G(-z_3)\G(-z_3+\delta)\G(1+z_3)\G(1+z_3-\delta) \right.\\
\left. \times \le -2\psi(1) + \psi(1+z_3) + \psi(1+z_3-\delta)+ \ln x\ri \right.\\ 
- \left. \delta \oint_C dz_2~dz_3~ \frac{x^{z_2}~y^{z_3}}{[23]^2} \G^2 \le -z_3 \ri  {\G^*} \le -z_2 -1 \ri  
\G \le -z_2 \ri  \G^2 \le z_3 + z_2 + 1 \ri + o(\delta)\right]  \\
= \frac{[23]-[31]-[12]}{2[31][23]^{1+\delta}} \G(1 - \delta)\G(1 + \delta)J(1,1-\delta,1) \\ 
+ \frac{(\ln x -2\psi(1))(1-y)}{2[23][31]}\oint_C~dz_3~y^{z_3}\G(-z_3)\G(-z_3+\delta)\G(1+z_3)\G(1+z_3-\delta)  \\
+ \frac{1-y}{2[23][31]}\oint_C~dz_3~y^{z_3}\G(-z_3)\G(-z_3+\delta)\G(1+z_3)\G(1+z_3-\delta) \\
\times \le\psi(1+z_3) + \psi(1+z_3-\delta) \ri \\ 
- \delta \oint_C dz_2~dz_3~ \frac{x^{z_2}~y^{z_3}}{[23]^2} \G^2 \le -z_3 \ri  {\G^*} \le -z_2 -1 \ri  
\G \le -z_2 \ri  \G^2 \le z_3 + z_2 + 1 \ri + o(\delta) \\
\equiv \frac{[23]-[31]-[12]}{2[31][23]^{1+\delta}} \G(1 - \delta)\G(1 + \delta)J(1,1-\delta,1) \\
+ \frac{(\ln x -2\psi(1))(1-y)}{2[23][31]} I'_4 + \frac{1-y}{2[23][31]} I'_5 - \frac{\delta}{[23]^2}I'_6 + o(\delta).  
\end{eqnarray*}
Integrals $I'_4$ and $I'_5$ are one-fold MB integrals in $z_3$ plane produced due to inclusion the residue at  
$z_2 = 0$ in $z_2$ plane in order to recover $J(1,1,1)$ construction. Again, we have added the residue instead of removing 
it due to clockwise orientation of $C.$ The one-fold MB integrals $I'_2,$ $I'_3,$ $I'_5,$  and two-fold  integral $I'_6$  
are defined as 
\begin{eqnarray*}
I'_2 = \oint_{C} dz_3~ y^{z_3} \left\{ \G\le -z_3 \ri \G(-z_3+\delta) \G \le z_3 + 1 \ri \G^* \le z_3 - \delta\ri \right\}, \\
I'_4 = \oint_C~dz_3~y^{z_3} \G(-z_3) \G(-z_3+\delta) \G(1+z_3) \G(1+z_3-\delta), \\
I'_5 =  \oint_C~dz_3~y^{z_3} \G(-z_3) \G(-z_3+\delta) \G(1+z_3) \G(1+z_3-\delta) \le \psi(1+z_3) + \psi(1+z_3-\delta) \ri,\\
I'_6 = \oint_C dz_2~dz_3~ x^{z_2}~y^{z_3} \G^2 \le -z_3 \ri  {\G^*} \le -z_2 -1 \ri \G \le -z_2 \ri  \G^2 \le z_3 + z_2 + 1 \ri
\end{eqnarray*}

\section[]{}
\setcounter{equation}{0}
\label{App:E}

In this Appendix we prove the invariance of the scalar triangle ladder integral with respect to Fourier transform via MB transform 
\footnote{This is a part of the talk of I.K. at the workshop ``High Energy Physics  in the LHC era,'' 
Valparaiso, Chile, 4-8 January 2010}. 
The explicit form of the UD functions $\Phi^{(n)}$ is given in Refs. \cite{Usyukina:1992jd,Usyukina:1993ch}
\begin{eqnarray}
\Phi^{(n)}\le x,y\ri = -\frac{1}{n!\L}\sum_{j=n}^{2n}\frac{(-1)^j j!\ln^{2n-j}{(y/x)}}{(j-n)!(2n-j)!}\left[{\rm Li_j}
\le-\frac{1}{\rho x} \ri - {\rm Li}_j(-\rho y)\right], \label{explicit}
\end{eqnarray}
\begin{eqnarray*}
\rho = \frac{2}{1-x-y+\L}, ~~~~ \L = \sqrt{(1-x-y)^2-4xy}.
\end{eqnarray*}
In Refs. \cite{Kondrashuk:2009us,Kondrashuk:2008ec} the following relations were derived for any $n>1.$
\begin{eqnarray*}
\frac{1}{[31]^2} \Phi^{(n)}\le \frac{[12]}{[31]},\frac{[23]}{[31]}\ri = 
\frac{1}{(2\pi)^4}\int~d^4p_1d^4p_2d^4p_3 ~ \delta(p_1 + p_2 + p_3) \times\no\\
\times e^{ip_2x_2} e^{ip_1x_1} e^{ip_3x_3} \frac{1}{(p_2^2)^2} \Phi^{(n)}\le \frac{p_1^2}{p_2^2},\frac{p_3^2}{p_2^2}\ri. 
\end{eqnarray*}
The simplest way to demonstrate these relations is to use the MB transform 
\begin{eqnarray*}
\Phi^{(n)}\le x,y\ri = \oint_C dz_2dz_3 x^{z_2} y^{z_3} {\cal M}^{(n)}\le z_2,z_3\ri
\end{eqnarray*}
with the MB-image ${\cal M}^{(n)}\le z_2,z_3\ri$ of the UD functions $\Phi^{(n)}\le x,y\ri.$ The precise form of 
${\cal M}^{(n)}\le z_2,z_3\ri$ can be found in Ref.~\cite{Usyukina:1993ch}. We do not need any use of the explicit
form of those MB images completely,
\begin{eqnarray*}
\frac{1}{(2\pi)^4}\int~d^4p_1d^4p_2d^4p_3 ~ \delta(p_1 + p_2 + p_3) 
e^{ip_2x_2} e^{ip_1x_1} e^{ip_3x_3} \frac{1}{(p_2^2)^2} \Phi^{(n)}\le \frac{p_1^2}{p_2^2},\frac{p_3^2}{p_2^2}\ri \\
= \frac{1}{(2\pi)^8}\int~d^4p_1d^4p_2d^4p_3d^4x_5 e^{ip_2(x_2-x_5)} e^{ip_1(x_1-x_5)} e^{ip_3(x_3-x_5)} 
\frac{1}{(p_2^2)^2} \Phi^{(n)}\le \frac{p_1^2}{p_2^2},\frac{p_3^2}{p_2^2}\ri \\
= \frac{1}{(2\pi)^8}\int d^4p_1d^4p_2d^4p_3d^4x_5\oint_C dz_2dz_3 ~ 
\frac{e^{ip_2(x_2-x_5)} e^{ip_1(x_1-x_5)} e^{ip_3(x_3-x_5)}}{(p_2^2)^{2+z_2+z_3} (p_1^2)^{-z_2} (p_3^2)^{-z_3}}  
{\cal M}^{(n)}\le z_2,z_3\ri  \\
= \frac{(4\pi)^6}{(2\pi)^8}\int d^4x_5 \oint_C dz_2dz_3 ~ 
\frac{\G(-z_2-z_3)}{\G(2+z_2+z_3)} \frac{\G(2+z_2)}{\G(-z_2)} \frac{\G(2+z_3)}{\G(-z_3)}\\ 
\times\frac{2^{2z_2+2z_3-2(2+z_2+z_3)}{\cal M}^{(n)}\le z_2,z_3\ri}{[25]^{-z_2-z_3}[15]^{2+z_2} [35]^{2+z_3} }  = \\
= \oint_C~dz_2dz_3 ~ \frac{{\cal M}^{(n)}\le z_2,z_3\ri}{[12]^{-z_3}[23]^{-z_2}[31]^{2+z_2+z_3} }  
= \frac{1}{[31]^2} \Phi^{(n)}\le \frac{[12]}{[31]},\frac{[23]}{[31]}\ri
\end{eqnarray*}

\end{appendix}

\end{document}